\documentclass[aps,pra,twocolumn,showpacs,amsmath,amssymb,preprintnumbers,superscriptaddress,10pt]{revtex4-1}

\usepackage{graphicx}
\graphicspath{{figures/}}
\usepackage{SIunits}
\usepackage{hyphenat}
\usepackage[bookmarks=false]{hyperref}
\usepackage{color}
\usepackage[capitalise]{cleveref}
\usepackage{array}
\usepackage{natbib}
\crefname{section}{Sec.}{Secs.}
\Crefname{section}{Section}{Sections}
\usepackage{makecell}

\definecolor{pink}{RGB}{255,0,255}

\definecolor{red}{rgb}{1,0,0}

\definecolor{blue}{RGB}{0,174,179}

\definecolor{vcc}{RGB}{0,0,179}

\begin{document}

\title{Controlling single-photon detector ID210 with bright light} 

\author{Vladimir~Chistiakov}
\email{v\_chistyakov@itmo.ru}
\affiliation{Faculty of Photonics and Optical Information, ITMO University, St.~Petersburg, Russia}

\author{Anqi~Huang}
\affiliation{Institute for Quantum Information \& State Key Laboratory of High Performance Computing, College of Computer, National University of Defense Technology, Changsha 410073, People's Republic of China}
\affiliation{Institute for Quantum Computing, University of Waterloo, Waterloo, ON, N2L~3G1 Canada}
\affiliation{\mbox{Department of Electrical and Computer Engineering, University of Waterloo, Waterloo, ON, N2L~3G1 Canada}}

\author{Vladimir~Egorov} 
\affiliation{Faculty of Photonics and Optical Information, ITMO University, St.~Petersburg, Russia}

\author{Vadim~Makarov}
\affiliation{Russian Quantum Center, Skolkovo, Moscow 143025, Russia}
\affiliation{\mbox{Shanghai Branch, National Laboratory for Physical Sciences at Microscale and CAS Center for Excellence in} \mbox{Quantum Information, University of Science and Technology of China, Shanghai 201315, People's Republic of China}}
\affiliation{NTI Center for Quantum Communications, National University of Science and Technology MISiS, Moscow 119049, Russia}
\affiliation{Department of Physics and Astronomy, University of Waterloo, Waterloo, ON, N2L~3G1 Canada}

\date{22 May 2019}

\begin{abstract}
We experimentally demonstrate that a single-photon detector ID210 commercially available from ID~Quantique is vulnerable to blinding and can be fully controlled by bright illumination. In quantum key distribution, this vulnerability can be exploited by an eavesdropper to perform a faked-state attack giving her full knowledge of the key without being noticed. We consider the attack on standard BB84 protocol and a subcarrier-wave scheme, and outline a possible countermeasure.
\end{abstract}

\maketitle

\section{Introduction}
\label{sec:intro}

Quantum key distribution (QKD) technology allows to securely distribute symmetric keys between two parties by utilizing fundamental aspects of quantum physics~\cite{bennett1984}. In theory, legitimate users (Alice and Bob) are able to detect any eavesdropping in the quantum channel performed by Eve. Today security of several QKD protocols has been unconditionally proven~\cite{renner2008}. However, in practice Eve is still able to obtain information about quantum keys without alarming Alice and Bob by exploiting loopholes in QKD hardware, which are not taken into consideration during the security analysis. This technique is referred to as ``quantum hacking'' and has been experimentally demonstrated with a variety of QKD components~\cite{lamas-linares2007,nauerth2009,xu2010,lydersen2010a,sun2011,jain2011,tang2013,bugge2014,sajeed2015,huang2016,makarov2016,huang2018}. These results have helped to further solidify QKD security by patching the loopholes or extending security analysis. It is therefore important to continue testing other QKD devices in order to develop efficient hacking countermeasures. \\

A particular QKD component found to be vulnerable to quantum hacking is a single-photon detector~\cite{makarov2006,lydersen2010a,lydersen2010b,wiechers2011,weier2011,lydersen2011b,lydersen2011c,sauge2011,gerhardt2011,sajeed2015a,huang2016}. For field applications in urban infrastructure, where the QKD nodes are located at medium distances (up to $100~\kilo\meter$), it is most practical to use single-photon registration systems based on avalanche photodiodes (APDs)~\cite{hadfield2009} because they provide sufficient efficiency without use of complex cooling systems required for superconducting detectors~\cite{natarajan2012}. Several experiments demonstrated that Eve can take full control over the detector by blinding it by an intense continuous wave (c.w.)\ laser and then sending additional trigger pulses in order to achieve controllable clicks at desired times. Combining this method with measuring photon states send by Alice allows Eve to secretly obtain full knowledge about the quantum key~\cite{gerhardt2011}. This quantum hacking technique is known as a faked-state attack \cite{makarov2005,lydersen2010a,gerhardt2011}. It has been implemented on several commercially available APDs~\cite{lydersen2010a,lydersen2010b,sauge2011,jogenfors2015,huang2016}.

The purpose of this work is to investigate the vulnerability to the faked-state attack of another single-photon detector, ID~Quantique ID210, which is currently commercially available \cite{idq-id210} and has recently been used in several QKD setups~\cite{canas2017,bannik2017,miroshnichenko2018}. Notably some of these experimental schemes are based on subcarrier-wave (SCW) QKD architecture where quantum states are formed at spectral sidebands of an intense light through phase modulation~\cite{gleim2017}. In SCW QKD systems a major fraction of the signal is filtered out before detection. Therefore another important task is to calculate realistic blinding parameters for SCW systems with ID~Quantique~ID210 detector. We have found that these setups are potentially susceptible to the faked-state attack. 

\section{Experimental setup}
\label{sec:setup}

In our tests we have used ID210 single-photon detector by ID~Quantique based on InGaAs/InP APD (unit serial number 1119019J010) \cite{idq-id210}. To simulate realistic conditions for Eve's attack, we have treated the detector as a black box in the course of all experiments. We have not opened its housing nor manually interfered in operation of any internal circuits. All APD settings have been at the values normally used in SCW QKD operation \cite{bannik2017}: quantum efficiency 10\%, gating frequency $100~\mega\hertz$, gate width $3~\nano\second$, deadtime $100~\nano\second$. For these settings, the dark count rate fluctuates around $200~\hertz$. All the parameters have been set using standard ID210 user interface from the front panel of the device.

\begin{figure}
	\includegraphics{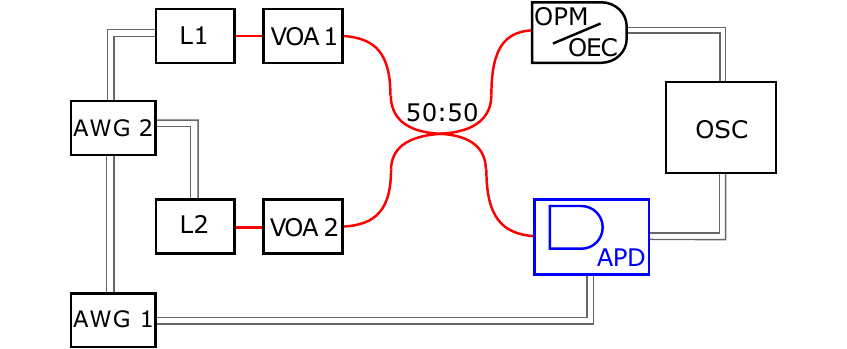}
	\caption{Experimental setup for testing the detector. L, laser; AWG, arbitrary waveform generator; VOA, variable optical attenuator; OPM, optical power meter; OEC, optical-to-electrical converter; OSC, oscilloscope; APD, avalanche photodiode single-photon detector ID~Quantique ID210.}
	\label{fig:exp-setup}
\end{figure}

Experimental setup for testing the detector for control by bright light is shown in~\cref{fig:exp-setup}. The APD is externally gated by an arbitrary waveform generator (AWG~1; Agilent 81110A) at frequency of $100~\mega\hertz$. This value is typical for SCW~QKD schemes~\cite{bannik2017,gleim2017}. Another generator (AWG~2; Highland Technology P400) is synchronized from AWG~1 and performs two functions. Firstly, it provides constant current to a continuous laser source (L1; Alcatel 1905 LMI) used for APD blinding. Secondly, it is driving the trigger pulse laser (L2; Gooch \& Housego AA1401) at $10~\mega\hertz$ rate. This value is lower than the gating frequency, because in realistic conditions only a small fraction of pulses emitted by Alice (no more than 10\%) reach Bob's single photon detector. L1 and L2 outputs are connected to variable optical attenuators (VOA~1; OZ Optics DA-100-3S-1550 and VOA~2; FOD 5418) that regulated output optical power. VOA outputs lead to fiber-optic beam splitter with a $50\!:\!50$ ratio. One beamsplitter output arm is connected to an optical power meter (OPM; Joinwit JW3208), while the other leads to the ID210 detector (APD). The power meter monitors optical power applied to the APD from L1 and L2. We have taken into account a non-ideal beamsplitting ratio when calibrating this power. At the second stage of experiment, we have substituted OPM with an optical-to-electrical converter limited to roughly $2~\giga\hertz$ bandwidth (OEC; LeCroy OE555), in order to accurately determine an optical pulse shape of L2. The electrical signals from OEC and APD are measured by an oscilloscope (OSC; LeCroy 820Zi). Trigger pulse energy is calculated from average optical power registered by OPM divided by the pulse repetition rate.

\section{Results}
\label{sec:results}

Our first task has been to find out if ID210 is susceptible to blinding. To do this, we have used L1 to generate c.w.\ laser radiation directed to APD optical input (with L2 switched off). L1 optical power has been regulated by VOA~1. When optical power at APD input has exceeded 24 nW, we have registered a complete absence of dark counts that indicates successful blinding of ID210 detector by c.w.\ radiation. The blinding does not harm the detector in any way, as its parameters return to normal each time L1 is turned off.

\begin{figure}
	\includegraphics{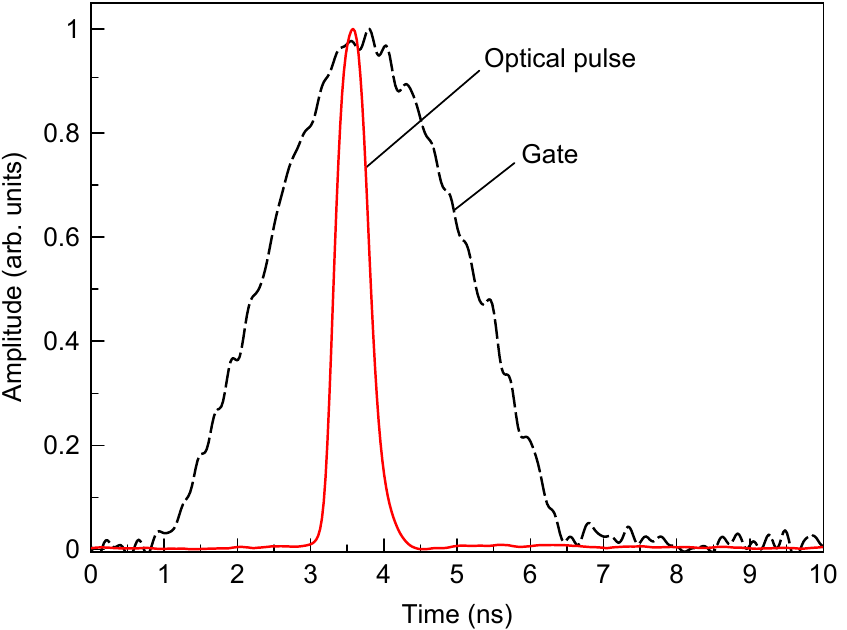}
	\caption{Oscillograms of the APD gate signal (provided at ID210 front-panel output) and the optical trigger pulse. Their relative timing is shown here as an assumption. The optical pulse width shown is limited by the bandwidth of OEC, i.e.,\ the actual pulse is shorter.} 
	\label{fig:pulse-shapes}
\end{figure}

Blinding the APD implies switching it from Geiger to linear mode by bright illumination. After that Eve can take full control over detector clicks by exceeding a current threshold at a comparator in the linear mode by sending trigger pulses of sufficient energy along with c.w.\ blinding radiation \cite{lydersen2010a}. Therefore our second step has been to determine the necessary trigger pulse parameters and synchronize these pulses with APD gates. 

We have initiated trigger pulses by turning on L2 with $5~\nano\second$ wide pulses at 10~MHz frequency. The latter value has been chosen as a maximum expected detector click frequency given $100~\nano\second$ deadtime. The shape of the optical trigger pulse is important for accurately adjusting the delay between the ``faked state'' and the detector gate (\cref{fig:pulse-shapes}). Its measured duration is less than $500~\pico\second$ full-width at half-magnitude (FWHM). Meanwhile, FWHM of the gate pulse matches the preset gate width of $3~\nano\second$. We have then adjusted the timing of our optical trigger pulse to minimize its energy required to produce a click in the blinded regime, which presumably aligns it with the middle of the gate.

\begin{figure}
	\includegraphics{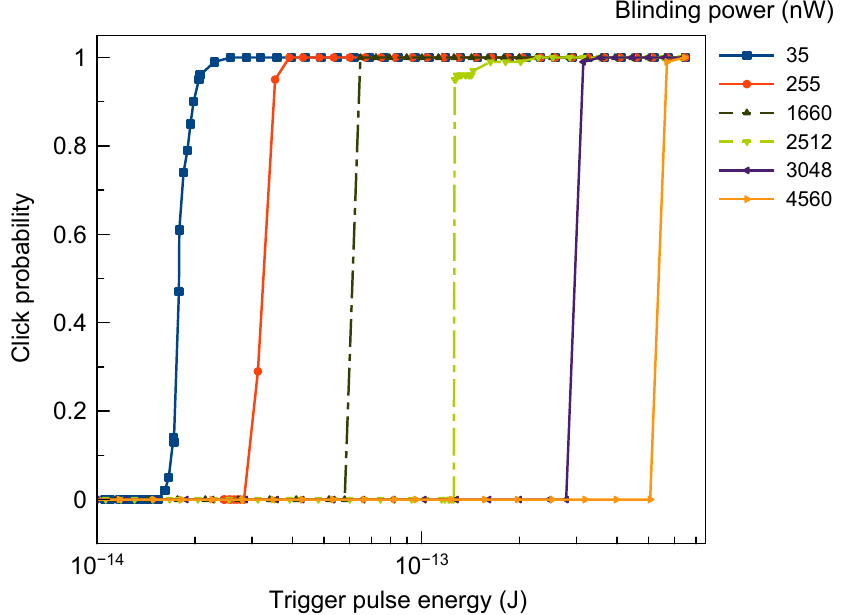}
	\caption{Detector click probability in the blinded regime as a function of control pulse energy.}
	\label{fig:detresponse}
\end{figure}

Our next step has been to determine a maximum trigger pulse energy $E_\text{never}$ at which the blinded detector still clicks with zero probability, and minimum energy $E_\text{always} $ at which it clicks with unity probability. When the trigger pulse energy $E_\text{trigger}$ is increased, the click probability undergoes a transition, shown in \cref{fig:detresponse} for several blinding powers. For example, under $35~\nano\watt$ c.w.\ blinding, the detector never clicks when $E_\text{trigger} \leq E_\text{never} = 15.4~\femto\joule$ and always clicks when $E_\text{trigger} \geq E_\text{always} = 25.8~\femto\joule$. At higher blinding powers, a click probability transition from 0 to 1 becomes more abrupt, which is apparent by comparing the plots for c.w.\ blinding of $35$ and $2512~\nano\watt$ that have been measured with a higher resolution to illustrate this effect.

The last step of characterizing the APD is defining the boundary values for trigger pulse energies that Eve can use to carry out the most efficient faked-state attack. Reference~\onlinecite{huang2016} describes in detail the methodology we use here for estimating these values. Let us consider BB84 protocol with four states in two orthogonal bases~\cite{bennett1984}. When Eve performs the faked-state attack, there are two possible detection outcomes: either Eve and Bob choose the same bases, or not. Eve wants Bob's detector to always click in the first case, and never in the second. She can achieve it by imposing a limitation on her $E_\text{trigger}$, making it sufficient to induce a click only when Bob's basis choice is the same as hers \cite{lydersen2010a}:
\begin{equation}
E_\text{always} \leq E_\text{trigger} \leq 2E_\text{never}.
\end{equation} 

\begin{figure}
	\includegraphics{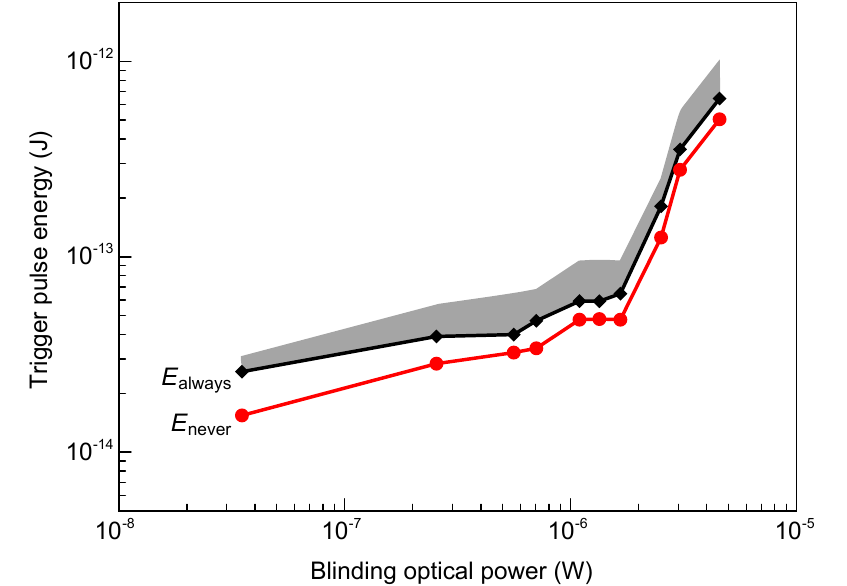}
	\caption{Click thresholds of investigated ID210 detector under different blinding powers. $E_\text{always}$ is a minimum pulse energy at which the detector always clicks and $E_\text{never}$ is a maximum energy when it never clicks. Shaded area shows the range of trigger pulse energies $E_\text{trigger}$ of the perfect attack (limited above by $2 E_\text{never}$).}
	\label{fig:blindcond}
\end{figure}

\Cref{fig:blindcond}~illustrates these boundaries for the analyzed ID210 detector: any trigger pulse energy between $E_\text{always}$ and $2E_\text{never}$, indicated by a shaded area, can be utilized for a successful attack. When Eve uses $E_\text{trigger}$ values from this interval, and Bob chooses the same basis, the eavesdropper will fully control the APD response and possess information on every key bit. When their bases do not coincide, a click will never happen, and these instances will be discarded by Alice and Bob during sifting stage. Thanks to this approach, Eve imposes on Bob only the states that she knows, and acquires full information about the quantum key. Hence, we have shown that an eavesdropper can perform a successful faked-state attack on ID210 single-photon detector in a realistic scenario of BB84 protocol.

\section{Attack on subcarrier-wave QKD}
\label{sec:scwqkd}

\begin{figure}
	\includegraphics{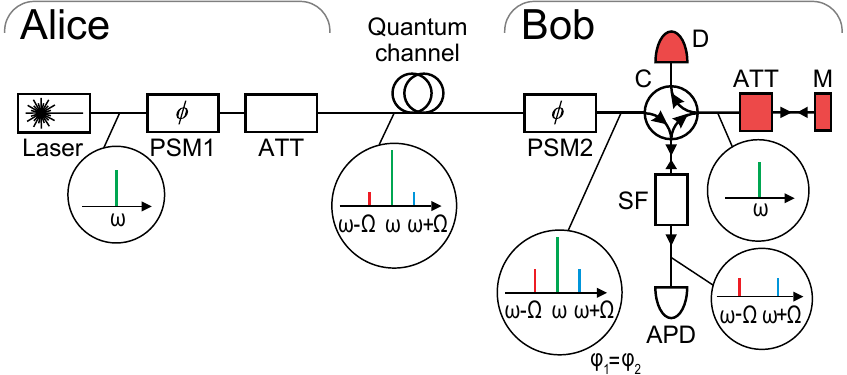}
	\caption{Subcarrier-wave QKD scheme. Components shaded red (gray) are introduced as a countermeasure, as described in \cref{sec:countermeasures}. PSM, electro-optical phase shift modulator; ATT, optical attenuator; C, circulator; SF, spectral filter; APD,  avalanche photodiode;  M, fiber-optic mirror; D, photodetector. Insets show optical spectra at different points.}
	\label{fig:scw-setup}
\end{figure}

The investigated detector has recently been employed in several QKD experiments~\cite{bannik2017,miroshnichenko2018} based on SCW principle~\cite{merolla1999}. This QKD scheme is promising as a backbone for large-scale quantum network thanks to its high capability for multiplexing~\cite{mora2012,chistiakov2014} and robustness against environmental influence on the optical fiber~\cite{gleim2017,gleim2016,kynev2017}. In this type of systems the encoding photons are not directly generated by an attenuated laser source but rather appear on spectral sidebands during a phase modulation of light. As can be seen from a general scheme of SCW QKD setup (\cref{fig:scw-setup}), the signal spectrum passes through a narrow filter (SF) before detection in order to remove the optical carrier that contains most of the optical power. It is therefore important to investigate if this filtering is an obstacle for Eve's detector control and faked-state attack on SCW QKD setups. In realistic conditions we should also consider insertion losses in Bob.

For our analysis we have used SCW QKD experimental parameters from Ref.~\onlinecite{gleim2017}: loss in Bob module $6.4~\deci\bel$, SF extinction ratio $30~\deci\bel$, and modulation index (the ratio between optical power in the carrier and the two sidebands) of~20. Let Eve prepare the signal states in a way similar to Alice: a spectrum with a strong carrier and two subcarriers. This spectrum will pass through the receiving unit undergoing the same modulation and filtering as the normal Alice's signal. Knowing the subcarrier power levels sufficient to blind the detector, we can estimate the total power that Eve should send into Bob module for successful blinding. For instance, let us consider the lowest blinding power of $35~\nano\watt$ confirmed experimentally in this work. In the SCW QKD scheme, before the pulse reaches the APD, it must undergo phase modulation at PSM2, where only $1/20$ of the initial optical carrier power is directed into the sidebands that will subsequently pass the SF. Therefore initial power at PSM2 input should be at least $700~\nano\watt$. Likewise, we should consider insertion loss in Bob's module ($6.4~\deci\bel$), therefore the minimum power used by Eve for a successful attack should be at least $3056~\nano\watt$. A~similar logic works for the trigger pulse energy, as summarized in~\cref{tab:blinding}. As can be seen, although Eve must operate with higher blinding powers and trigger pulse energies in order to control the detector in SCW QKD scheme, the power levels needed are still sufficiently low not to damage any optical components \cite{makarov2016}. These results suggest that SCW QKD setups do not have enough intrinsic loss to prevent detector control using the described method.

\begin{table}
	\caption{\label{tab:blinding}Calculated parameters for successful control of ID210 detector in SCW QKD scheme.}
	\begin{tabular}[t]{c c c c}
	\hline\hline
	\makecell{Eve's faked-state\\ power in...} & \makecell{Blinding\\ power ($\nano\watt$)} & $E_\text{always}$ ($\femto\joule$) & $E_\text{never}$ ($\femto\joule$) \\
	\hline
	\makecell{subcarriers after\\ filtering} & 35 & 25.8 & 15.4 \\
	\makecell{spectrum before\\ modulation} & 700 & 516 & 308 \\
	\makecell{spectrum entering\\ Bob's module} & 3056 & 2252 & 1345 \\
	\hline\hline
	\end{tabular}
\end{table}

\section{Countermeasures}
\label{sec:countermeasures}

The faked-state attack is very general and has been successfully used for hacking different APDs. The most efficient countermeasure against it is implementing measurement-device-independent~(MDI) QKD architecture~\cite{lo2012}, where the detection unit is moved from Alice and Bob to an untrusted party Charlie. MDI~QKD protocol is based on Bell state measurements and ensures that Charlie (or Eve) is limited to openly announcing the measurement outcomes and is incapable of acquiring secure key information. Unfortunately, in practice MDI~QKD architecture remains difficult to implement and yields much lower key rates.

In traditional two-party QKD, attempts to produce a countermeasure of a similar quality integrated with a security proof have led to stringent requirements on components \cite{lydersen2011a,maroy2017}, which have not yet been implemented. Simpler countermeasures that utilize photon counting statistics have been proposed but none yet battle-tested \cite{honjo2013,koehler-sidki2018a,fedorov2019}. A more practical countermeasure may imply redesigning an avalanche quenching circuit of the APD and introducing precise photocurrent sensors into it \cite{yuan2010,lydersen2010c,yuan2011,lydersen2011d,yuan2011a,koehler-sidki2018}.

Here we propose a simple solution for the SCW~QKD scheme analyzed in \cref{sec:scwqkd}. As explained above, in SCW~QKD intense optical radiation acts as a carrier for the phase-modulated quantum signal on its sidebands. Even though the carrier contains no information about the key, in practical QKD it is necessary to detect it as a countermeasure against a photon-number-splitting attack~\cite{guerreau2005}. We propose to reveal APD blinding by monitoring this signal. Our system contains a circulator used to measure the carrier and sideband signals individually (\cref{fig:scw-setup}). Since the faked-state attack requires significantly elevated carrier optical power (see~\cref{tab:blinding}), a watchdog detector D can be installed for monitoring its abnormally high values. We presume that one cannot put an unprotected detector into a third port of the circulator, as it could be potentially blinded by Eve there. We therefore suggest to place it in a fourth port and protect it by an attenuator and a mirror in the third port, as shown in~\cref{fig:scw-setup}. The attenuation value should be carefully chosen to be high enough to prevent blinding of D but sufficiently low to allow carrier detection by a regular photodiode. Testing this idea can be future work.

\section{Conclusion}
\label{sec:conclusion}

We have demonstrated experimentally that ID~Quantique ID210 single-photon detector based on avalanche photodiode is vulnerable to blinding and can be controlled by bright light. We have shown that the faked-state attack will work in SCW~QKD systems where a major signal fraction is filtered out before detection. We have also suggested a simple optical scheme that could act as a potential countermeasure in  SCW~QKD. Overall, even though the faked-state attack was introduced a decade ago, no universal industrial-scale solution for two-party QKD has been found yet. Today MDI~QKD remains the only strictly proven countermeasure against detector hacking. All alternative solutions are still to be meticulously tested and incorporated into existing security proofs. Our results emphasise that known vulnerabilities should be addressed at the system design stage, and any countermeasures thoroughly tested experimentally.

\acknowledgments
This Letter has been reviewed by ID~Quantique prior to its publication. This work was funded by the Ministry of Education and Science of Russia (programs 5-in-100 and NTI center for quantum communications), NSERC of Canada (programs Discovery and CryptoWorks21), CFI, and MRIS of Ontario. A.H.\ was supported by China Scholarship Council. This work was funded by Government of Russian Federation (grant 08-08).

{\em Author contributions:} V.C.\ conducted the experiment, analyzed data, proposed the countermeasure, and wrote the Letter with input from all authors. A.H.\ guided the experiment. V.M.\ and V.E.\ supervised the study.

\def\bibsection{\medskip\begin{center}\rule{0.5\columnwidth}{.8pt}\end{center}\medskip} 
\bibliography{library}

\end{document}